\documentstyle[aps,preprint]{revtex}
\begin{document}
\draft
\title{Dependence of the flux creep activation energy on current density and magnetic field for MgB$_2$ superconductor}

\author{M. J. Qin, X. L. Wang, H. K. Liu, S. X. Dou}

\address{Institute for Superconducting \& Electronic Materials, University of Wollongong, Wollongong, NSW 2522, Australia}

\maketitle

\begin{abstract}
Systematic ac susceptibility measurements have been performed on a MgB$_2$ bulk sample. We demonstrate that the flux creep activation energy is a nonlinear function of the current density $U(j)\propto j^{-0.2}$, indicating a nonlogarithmic relaxation of the current density in this material. The dependence of the activation energy on the magnetic field is determined to be a power law $U(B)\propto B^{-1.33}$, showing a steep decline in the activation energy with the magnetic field, which accounts for the steep drop in the critical current density with magnetic field that is observed in MgB$_2$. The irreversibility field is also found to be rather low, therefore, the pinning properties of this new material will need to be enhanced for practical applications. 

\end{abstract}
\vspace{1cm}


The recent discovery of the new superconductor MgB$_2$ \cite{naga} with its transition temperature at 39K has aroused considerable interest in the field of condensed matter physics, especially in the area of superconductivity. Intensive studies have been carried out on this new material by means of magnetic \cite{larb,taka,bugo,chen,wenh,lima,josh,finn} and transport \cite{taka,finn,budk,kang,kijo} measurements, microstructure studies \cite{sung}, and other experimental techniques \cite{kang1,kara,schm,takah,jorg}, as well as theoretical calculations \cite{medv,voel,shul}. The fundamental superconducting parameters of MgB$_2$ such as the upper critical field $B_{c2}(0)=12.5-18$ T \cite{larb,taka,wenh,josh,budk}, the lower critical field $B_{c1}(0)\approx 300$ Gs \cite{taka,josh}, the Ginzburg-Landau parameter $\kappa\approx 26$ \cite{finn}, the coherence length $\xi(0)=4-5.2$ nm \cite{larb,finn}, the penetration length $\lambda (0)=140-180$ nm \cite{chen,finn}, and the energy gap $\Delta(4.2 {\rm K})=4.3-5.2$ meV \cite{kara,schm} have been obtained. The critical current density, one of the most important parameters in considering superconductors for practical applications, has been determined to be about $10^6$ A/cm$^2$ at 0 T and 4.2 K in bulk samples \cite{wenh} and $8\times 10^6$ A/cm$^2$ at 0 T and 5 K in thin films \cite{kangw}. The critical current density is determined by the pinning properties of the sample as well as by the flux motion, because the motion of the vortices over pinning centers (flux creep) in the superconductor induces dissipation and reduces the critical current density $j_c$. It is the flux creep that sets the limiting critical current density in superconductors. It is thus essential to study the activation energy against flux motion, in order to understand the underlying mechanism and therefore to enhance the current carrying capacity of this new material. In this paper, we investigate the flux creep activation energy in MgB$_2$, and determine its dependence on the current density, the magnetic field, and the temperature by measuring the real $\chi'(T)$ and imaginary $\chi''(T)$ parts of the ac susceptibility at different ac field amplitudes, frequencies and dc magnetic fields.

All measurements have been performed on a MgB$_2$ bulk sample, which was prepared by conventional solid state reaction \cite{dous}. High purity Mg and B (amorphous) were mixed and finely ground, then pressed into pellets 10 mm in diameter with 1-2 mm thickness. Extra Mg was added in order to avoid Mg loss at high temperatures. These pellets were placed on an iron plate and covered with iron foil, then put into a tube furnace. The samples were sintered at temperatures between 700 and 1000$^o$C for 1-14h. A high purity Ar gas flow was maintained throughout the sintering process. A sample with dimension of $2.18\times 2.76\times 1.88$ mm$^3$ was cut from the pellet. Phase purity was determined by XRD and grain size by SEM. Only a small level of MgO was found and the grain size was determined to be about 200 $\mu$m. 

The ac susceptibility measurements were carried out using a superconducting quantum interference device (SQUID, Quantum Design PPMS 8 T). The superconducting transition temperature $T_c=38.6$ K, $\Delta T_c<1$ K was determined by ac susceptibility in an ac field of 0.1 Gs and frequency 117 Hz.

Fig. \ref{fig1} shows the effects of the dc magnetic field $B_{dc}$ on the ac susceptibility of the MgB$_2$ bulk sample. As $B_{dc}$ is increased from 0.5 T to 3 T, the transition temperature shifts to lower temperatures and the transition width broadens. Although the transition width is slightly changed from about 3 K at 0.5 T to 6 K at 3 T, the transition temperature is greatly depressed by the dc field from about 35 K at 0.5 T to 27 K at 3 T, while in YBCO, the depression in $T_c$ is quite small \cite{ding,qin1}. As the transition temperature under a dc magnetic field is an indication of the irreversibility line, this result indicates that the irreversibility line in the MgB$_2$ bulk sample is rather low in the H-T plane, similar to the results obtained by dc magnetization measurements \cite{larb,wenh}.

Shown in Fig. \ref{fig2} are typical $\chi'(T)$ and $\chi''(T)$ curves for the MgB$_2$ bulk sample at $B_{dc}$=1 T, $f=1117$ Hz and different ac field amplitudes $B_{ac}$ indicated. As $B_{ac}$ is increased, the transition shifts to lower temperatures with increased width.
In Fig. \ref{fig3}, we show the effects of the frequency on the ac susceptibility of this MgB$_2$ bulk sample. Contrary to the effects of $B_{ac}$ and $B_{dc}$, as $f$ is increased, the transition shifts to higher temperatures and the transition width broadens. 
All the characteristics shown in Figs. \ref{fig1},\ref{fig2},\ref{fig3} for the MgB$_2$ bulk sample are similar to what have been observed in high temperature superconductors \cite{ding,qin1} and predicted from theoretical calculations \cite{qin2}. This is understandable, because ac susceptibility at different dc magnetic fields, ac field amplitudes and frequencies reflects a common phenomenon, i.e., flux dynamics in type-II superconductors.

A measurement of the superconducting transition by means of the ac susceptibility $\chi = \chi'+i\chi''$  typically shows, just below the critical temperature $T_c$, a sharp decrease in the real part of the susceptibility $\chi'$, a consequence of diamagnetic shielding, and a peak in the imaginary part of the susceptibility $\chi''$, representing losses.
The peak in $\chi''$ will occur when the flux front reaches the center of the sample. It follows that the position of the peak in
$\chi''$ will also strongly depend on temperature, dc field, ac field amplitude and frequency. The criterion for the peak in $\chi''$ is \cite{blat}
\begin{equation}
U(T_p, B_{dc}, j) = k_{\rm B}T_p{\displaystyle \ln\frac {1}{f_{\rm peak}t_0}}
\label{eq1}
\end{equation}
where the time scale $t_0=4\pi\mu_0H_{ac}^2/\rho_0j^2(\omega)$ \cite{blat}, $\rho_0$ is the prefactor in the Arrhenius law $\rho = \rho_0 \exp [- U(j)/k_{\rm B}T]$, $T_p$ is the peak temperature in the $\chi''(T)$ curve and $k_{\rm B}$ the Boltzmann constant.

It has been shown \cite{qin2} by numerical calculation that during the penetration by the ac magnetic field into a superconductor, the magnetic field profile can be regarded as a straight line, so at the peak temperature the current density can be approximated as 
\begin{equation}
j = {\displaystyle \frac {H_{ac}}{d}}
\label{eq2}
\end{equation}
where $d$ is the sample size.

A plot of $-\ln f_{\rm peak}$ versus $U(T_p)/k_{\rm B}T_p$ should thus be a straight line with the slope of $U(j,B_{dc})$,
\begin{equation}
{\displaystyle \frac {U(T)}{k_{\rm B}T}} U(j, B_{dc})  = -\ln (f) - \ln (t_0)
\label{eq3}
\end{equation}
By varying the ac amplitude and then using Eq.(\ref{eq2}) to determine the current density, one can then reconstruct the current density dependence of the activation energy $U(j)$. Using the ac method the usual difficulty in conventional relaxation measurements of having only a very limited time window ($1 \sim 10^4$ s) can be overcome by extending the latter to smaller values of $10^{-5} \sim 10^{-3}$ s ($f=100$ kHz $\sim $ 1 kHz) \cite{ding,qin1}.

In order to account for the explicit temperature dependence of the activation energy, we choose a form of temperature scaling function
\begin{equation}
U(T)=[1-(T/T_x)^2]^2
\label{eq4}
\end{equation}
where $T_x=36.3, 34.3, 31.5, 29.1$ K for $B_{dc}=0.5, 1, 2, 3$ T respectively is a  characteristic temperature, which is taken from the
magnetic irreversibility line. $U(T)$ changes slightly with
temperature for $T\ll T_x$ and drops rapidly as T approaches $T_x$. A detailed
discussion on choosing the function $U(T)$ has been given by McHenry et al \cite{mche}.

Fig. \ref{fig4} shows $-\ln f_{\rm peak}$ versus $U(T_p)/k_{\rm B}T_p$ curves at $B_{dc}=0.5$ T and various current densities. The experimental data can be fitted very well by straight lines (Eq.(\ref{eq3}), solid lines in Fig.\ref{fig4}). We can then derive the activation energy $U(j,B_{dc}=0.5 {\rm T})$ from the slopes of the straight lines. $U(j,B_{dc})$ at other dc magnetic fields have also been derived, and the results are summarized in Fig. \ref{fig5}, where the activation energy $U(j)\propto U(j,B_{dc})\times B^{1.3}$ is plotted as a function of the current density for the MgB$_2$ bulk sample at various dc magnetic fields. As can be seen from Fig. \ref{fig5}, we have obtained a universal curve $U(j)$ by scaling the data by $B^{1.3}$. The slight scattering at low current density may result from the field dependent critical current density $j_c(B)$. Note that $B_{ac}$ has been changed to $j$ by using Eq.(\ref{eq2}), where $d$ is the sample size rather than the grain size, because it has been reported \cite{larb} that current flow in MgB$_2$ is strongly linked. The current density $j$ obtained here is also very closed to what has been derived using magnetization measurements \cite{dous}.

From the best fit of the data in Fig. \ref{fig5}, we derived the current density dependent activation energy $U(j)\propto j^{-0.2}$, which is highly nonlinear. This result suggests that the I-V curve of MgB$_2$ should also be highly nonlinear, because using the Arrhenius rate equation, we have $E=Bv_0=Bv_0\exp(-U(j)/k_{\rm B}T)\propto \exp(-j^{-\mu})$. Nonlinear $I-V$ characteristics have been experimentally observed in MgB$_2$ \cite{kijo}. On the other hand, the relaxation of the current density or the magnetization can be derived from Eq.(\ref{eq1}) as $j(t)\propto [\ln(t/t_0)]^{-1/\mu}$, which is also a nonlinear function of $\ln(t/t_0)$. This nonlogarithmic relaxation can be detected by means of dc magnetization relaxation measurements, which will be discussed in one of our forthcoming works.

As can be seen from Eq.(\ref{eq3}), with the current density $j$ fixed, we can also derive the activation energy as a function of the dc magnetic field $U(B)$. The results are summarized in Fig. \ref{fig6}, where the activation energy $U(B)\propto U(j,B_{dc})\times j^{0.21}$ is plotted as a function of the magnetic field for the MgB$_2$ bulk sample at various current densities. As can be seen from Fig. \ref{fig6}, we have also obtained a universal curve by scaling the data by $j^{0.21}$. This current density dependence is consistent with the one derived in Fig. \ref{fig5}. As the scaling factor $B_0$ (see Eq.(\ref{utbj}) below) for $B$ is current density independent, we can see that the scaling of $U(B)$ is much better than that of $U(j)$ shown in Fig. \ref{fig5}. The solid line in Fig. \ref{fig6} is a fit to the power law $U(B)\propto B^{-1.33}$. The obtained $U(B)$ is also consistent with the one derived from scaling in Fig. \ref{fig5}. The self-consistent scalings of $U(j,B)$ shown in Fig. \ref{fig5} and Fig. \ref{fig6} suggest that the separation of the activation energy $U(j,B,T)$ to $U(j)U(T)U(B)$ is quite reasonable. The final expression for the temperature, field and current density dependent activation energy is given by
\begin{equation}
U(T,B,j)=U_0\left [1-\left(\frac {T}{T_x}\right)^2\right ]^2\left(\frac {B}{B_0}\right)^{-n}\left(\frac {j_0}{j}\right)^\mu
\label{utbj}
\end{equation}
where $U_0$, $B_0$ and $j_0$ are scaling values, and the exponents $n$ and $\mu$ are determined to be 1.33 and 0.2 respectively.

As for the magnetic field dependence of the activation energy, a $B^{-1}$ dependence has been previously derived using the Anderson-Kim model of the activation energy combined with the Ginzburg-Landau expressions for the coherence length, thermodynamic critical field and depairing critical current density, etc. \cite{yesh,tink}. Such a $B^{-1}$ dependence has been observed in a La$_{1.86}$Sr$_{0.14}$CuO$_4$ single crystal with weak pinning centers by McHenry et al. \cite{mche}. And 
for YBa$_2$Cu$_3$O$_x$ samples with strong pinning centers, such as twin planes, stacking faults or Y$_2$BaCu$O_5$ inclusions,
a $U(B) \sim B^{-0.5}$ has been derived by both ac susceptibility \cite{qin1} and dc magnetization measurements \cite{kung1,kung2}. 

For the new superconductor MgB$_2$ on the other hand, we find a $U(B)\sim B^{-1.33}$ dependence, showing that the activation energy decreases even faster with increasing magnetic field, compared to weakly pinned high temperature superconducting La$_{1.86}$Sr$_{0.14}$CuO$_4$ single crystal. The weakening of the activation energy with increasing magnetic field may be the reason why the critical current density drops steeply as the magnetic field increases, as has been observed by dc magnetization measurements \cite{larb,taka,bugo,chen,wenh,finn}. Taking into account the relatively low irreversibility line (Fig. \ref{fig1}) and the weakening of the activation energy with increasing magnetic field, the pinning properties of MgB$_2$ need to be enhanced for practical applications.

In summary, we have performed systematically ac susceptibility measurements on a MgB$_2$ bulk sample. The magnetic field and current density dependent flux creep activation energy has been determined to be $U(j,B)\propto j^{-0.2}B^{-1.33}$. Compared to high temperature superconductors $U(B)\sim B^{-1}$ for weakly pinned La$_{1.86}$Sr$_{0.14}$CuO$_4$ single crystal and 
$U(B)\sim B^{-0.5}$ for strongly pinned YBa$_2$Cu$_3$O$_x$, the steeply declining dependence $U(B)\sim B^{-1.33}$ results in a steep drop in $j_c$ with magnetic field and suggests that pinning in MgB$_2$ is quite weak, as can also be seen from the low irreversibility field. 

The authors would like to thank Australian Research Council for financial support.

\begin{figure}
\caption{$\chi'$(T) and $\chi''$(T) curves of the MgB$_2$ bulk sample at $B_{ac}$=1 Gs, $f=1117$ Hz and $B_{dc}=0.5, 1, 2, 3$ T.}
\label{fig1}
\end{figure}

\begin{figure}
\caption{$\chi'$(T) and $\chi''$(T) curves of the MgB$_2$ bulk sample at $B_{dc}$=1 T, $f=1117$ Hz and $B_{ac}=0.1, 0.5, 1, 2, 5, 10, 15$ Gs (from right to left).}
\label{fig2}
\end{figure}

\begin{figure}
\caption{$\chi'$(T) and $\chi''$(T) curves of the MgB$_2$ bulk sample at $B_{dc}$=0.5 T, $B_{ac}=2$ Gs and $f=17, 51, 117, 351, 1117, 3331, 9999$ Hz (from left to right).}
\label{fig3}
\end{figure}

\begin{figure}
\caption{$-\ln f_{\rm peak}$ versus $U(T_p)/k_{\rm B}T_p$ of the MgB$_2$ sample at various current densities indicated by different symbols. Solid lines are linear fits calculated from Eq.(\ref{eq3}).}
\label{fig4}
\end{figure}

\begin{figure}
\caption{Activation energy $U(j)\propto U(j,B_{dc})\times B^{1.3}$ as a function of the current density for the MgB$_2$ bulk sample at various dc magnetic fields. Solid line is the fitting curve $U(j)\propto j^{-0.2}$.}
\label{fig5}
\end{figure}

\begin{figure}
\caption{Activation energy $U(B)\propto U(j,B_{dc})\times j^{0.21}$ as a function of the magnetic field for the MgB$_2$ bulk sample at various current densities. Solid line is the fitting curve $U(B)\propto B^{-1.33}$.}
\label{fig6}
\end{figure}

\end{document}